\documentclass[twocolumn]{aastex631}

\usepackage{physics}
\usepackage{color}
\usepackage{xspace}

\DeclareRobustCommand{\okina}{%
  \raisebox{\dimexpr\fontcharht\font`A-\height}{%
    \scalebox{0.8}{`}%
  }%
}

\begin{document}

\title{Fast Radio Bursts and Interstellar Objects}

\author[0000-0002-0924-8403]{Dang Pham}
\thanks{DP and MJH contributed equally.}
\affiliation{Department of Astronomy and Astrophysics, University of Toronto, Toronto, Ontario, M5S 3H4, Canada}

\author[0000-0001-6314-873X]{Matthew J. Hopkins}
\thanks{DP and MJH contributed equally.}
\affiliation{Department of Physics, University of Oxford, Denys Wilkinson Building, Keble Road, Oxford, OX1 3RH, UK}

\author[0000-0001-5578-359X]{Chris Lintott}
\affiliation{Department of Physics, University of Oxford, Denys Wilkinson Building, Keble Road, Oxford, OX1 3RH, UK}

\author[0000-0003-3257-4490]{Michele T. Bannister}
\affiliation{School of Physical and Chemical Sciences—Te Kura Mat\=u, University of Canterbury, Private Bag 4800, Christchurch 8140, Aotearoa
New Zealand}

\author[0000-0003-1927-731X]{Hanno Rein}
\affiliation{Department of Astronomy and Astrophysics, University of Toronto, Toronto, Ontario, M5S 3H4, Canada}
\affiliation{Department of Physical and Environmental Sciences, University of Toronto at Scarborough, Toronto, Ontario M1C 1A4, Canada}

\email{dang.pham@astro.utoronto.ca, matthew.hopkins@physics.ox.ac.uk}

\begin{abstract}
Fast radio bursts (FRBs) are transient radio events with millisecond-scale durations, and debated origins.
Collisions between planetesimals and neutron stars have been proposed as a mechanism to produce FRBs; the planetesimal strength, size and density determine the time duration and energy of the resulting event.
One source of planetesimals is the population of interstellar objects (ISOs), free-floating objects expected to be extremely abundant in galaxies across the Universe as products of planetary formation.
We explore using the ISO population as a reservoir of planetesimals for FRB production, finding that the expected ISO-neutron star collision rate is comparable with the observed FRB event rate.
Using a model linking the properties of planetesimals and the FRBs they produce, we further show that observed FRB durations are consistent with the sizes of known ISOs, and the FRB energy distribution is consistent with the observed size distributions of Solar System planetesimal populations.
Finally, we argue that the rate of ISO-neutron star collisions must increase with cosmic time, matching the observed evolution of the FRB rate.
Thus, ISO-neutron star collisions are a feasible mechanism for producing FRBs.
\end{abstract}


\section{Introduction}
\label{sec:intro}

Fast radio bursts (FRBs) are gigahertz radio-emitting, millisecond-timescale transient events. 
First detected in 2007 \citep{Lorimer_2007}, a large catalogue has been assembled by the Canadian Hydrogen Intensity Mapping Experiment \citep[CHIME;][]{CHIME/FRBCollaboration_2021}.
The majority of known FRBs are extragalactic \citep{Mannings_2021}, with only one yet known in the Milky Way \citep{Bochenek_2020,CHIME/FRBCollaboration_2020}.
They exist in a wide range of host galaxies and local environments, but remain sparse: the inferred volumetric FRB rate for events above a threshold energy of \(10^{35}\mathrm{~erg}\) 
is \((7^{+9}_{-6})\times10^{7}\mathrm{\, Gpc^{-3} \,yr^{-1}}\) \citep{Bochenek_2020}.
Two types are known: single (one-off) events, and repeating FRBs.

One of the many \citep[cf.][]{Platts_2019, ZhangB_2020} proposed mechanisms to produce an FRB is a collision between a planetesimal and a neutron star (NS), initially suggested by \cite{Geng_2015}.
Though neutron stars have been found to have planetary systems bound to them \citep{Wolszczan_1992}, the radiation mechanism of \cite{Geng_2015} is specifically for unbound planetesimals, colliding on radial trajectories, as bound planetesimals are expected to interact differently \citep{Brook_2014}.
\cite{Dai_2016} and \cite{Bagchi_2017} invoked the mechanism of \cite{Geng_2015} to explain repeating FRBs as NS undergoing frequent encounters in debris disks around other stars.
However, the feasibility of this scenario producing repeating FRBs at the rate observed is debated \citep{Smallwood_2019, Deng_2024}.
Nevertheless, non-repeating FRBs may still be generated by NS-planetesimal collisions, if impacts can occur at a suitable rate in free space.

In this work we consider the collisions between NS and interstellar objects (ISOs).
An expected feature of planet formation \citep{McGlynn_1989}, ISOs are planetesimals unbound from the planetary systems they formed in \citep{'OumuamuaISSITeam_2019} with a density of \(10^{15}~\mathrm{pc}^{-3}\) in the Solar neighbourhood \citep{Do_2018} and a total number of \(\sim10^{27}\) across the Milky Way, and two discovered so far: 1I/\okina Oumuamua and 2I/Borisov.
We first quantify the ISO-NS collision rate and consider its implications (Section \ref{sec:rate}).
We then use the radiation mechanism of \cite{Dai_2016} to compare the observed durations and energetics of FRBs to the sizes of ISOs (Section \ref{sec:properties}).
Finally in Section \ref{sec:discussion} we discuss testable implications of FRBs being produced by ISO-NS collisions, and challenges with scaling the ISO-NS collision rate with the FRB event rate.

\section{ISO-NS Collision Rate}
\label{sec:rate}
In this section we first detail our estimate for the ISO-NS collision rate, then discuss sources of uncertainty in our estimate.

To estimate the density and velocity distribution of ISOs around NS we make a number of assumptions.
Since ISOs are an abundant, expected product of planetary systems \citep{Shoemaker_1984,Stern_1990,Jewitt_2023}, we first assume that planetary systems and therefore ISO populations are present in galaxies throughout the Universe. 
The population of ISOs larger than 1I/\okina Oumuamua passing through the Solar system has a number density of \(n_\textrm{ISO}\sim10^{15} ~\mathrm{pc}^{-3}\) \citep{Do_2018}, and a velocity distribution with a width of approximately \(50\mathrm{~km~s^{-1}}\) \citep{Hopkins_2024}.
Being estimated from a single detection in a well-characterised survey this estimated number density has an uncertainty of an order of magnitude, but lacking additional constraints we assume the number density of ISOs around all NS will be the same order of magnitude as this.
For the relative velocities of NS and ISOs, we note that upon creation NS receive a large natal kick on the order of \(\sim100\,\mathrm{km~s^{-1}}\) \citep{Lyne_1994,Hobbs_2005}. 
As this is significantly larger than the ISO velocity dispersion around the Sun, we assume this value for the relative speed \(v_\infty\) of ISOs encountering NS.

In the vicinity of a NS of mass \(M_\mathrm{NS}\simeq1.4\mathrm{~M_\odot}\) \citep{Heger_2003}, ISOs follow hyperbolic trajectories defined by their impact parameter \(b\) and relative speed at infinity \(v_\infty\), related to the periastron \(q\) by \(b^2 = q^2 \left( 1 + \frac{2 G M_\mathrm{NS}}{q v_\infty^2} \right)\).
An ISO impacts the NS surface when it has a periastron less than the NS radius \(R_\mathrm{NS}\simeq 10\,\mathrm{km}\) \citep{Miller_2021}, therefore the cross-section of ISOs with a relative speed at infinity \(v_\infty\) colliding with the NS surface is 
\begin{equation}
\begin{split}
\sigma & = \pi b^2 =\pi R_\mathrm{NS}^2\left(1+\frac{2GM_\mathrm{NS}}{v_\infty^2 R_\mathrm{NS}}\right)\\
& \approx 10^{9} \mathrm{~km^2}\left(\frac{M_\mathrm{NS}}{1.4\mathrm{~M_\odot}}\right)\left(\frac{R_\mathrm{NS}}{10\mathrm{~km}}\right)\left(\frac{v_\infty}{100\mathrm{~km~s^{-1}}}\right)^{-2}\,.
\end{split}
\end{equation}
The increase in cross-section due to the second term is referred to as gravitational focussing \citep[cf.][]{Forbes_2019} and completely dominates in this high-gravity situation.
Thus, taking \(n_\text{ISO}\sim10^{15}\mathrm{~pc}^{-3}\), \(v_\infty\sim100\mathrm{~km~s}^{-1}\) and \(\sigma\sim10^{9} \mathrm{~km^2}\) gives the encounter rate of a single NS with ISOs as 
\begin{equation}\label{eq:gamma}
\Gamma =  n_\mathrm{ISO} v_\infty \sigma \sim 10^{-7}\,\mathrm{yr}^{-1}.
\end{equation}

FRBs are detected over cosmological distances, with the most distant being detected at \(z=1\) \citep{Ryder_2023}, thus we must consider the cosmological rate of ISO-NS collisions.
Stars over \(9M_\odot\) make up \(\sim 0.5\%\) of all stars formed \citep[assuming the universal initial mass function of][]{Kroupa_2001}, have lifetimes \(\lesssim 0.1\mathrm{~Gyr}\) and largely form neutron stars upon their deaths \citep{Heger_2003}.
Thus with approximately \(10^{11}\) stars \citep[e.g.][]{Licquia_2015}, the Milky Way contains about \(N_\mathrm{NS}\sim10^9\) NS. 
For a number density of Milky Way-mass galaxies of \(n_\mathrm{Gal}\sim10^{7}\,\mathrm{Gpc}^{-3}\) \citep{Blanton_2003}, the cosmological rate of ISO-NS collisions is
\begin{equation}
R_\mathrm{col} = \Gamma \cdot n_\mathrm{Gal} \cdot N_\mathrm{NS} \sim 10^9~\mathrm{Gpc^{-3}yr^{-1}}\,.
\end{equation}
FRBs produced by ISO-NS collisions are thought to be strongly beamed by a factor \(f\sim10^{-2}\) \citep{Dai_2020} meaning if every 1I-sized ISO-NS collision produced an FRB the observable rate would be
\begin{equation}\label{eqn:R_obs}
R_\mathrm{obs} = f R_\mathrm{col} \sim 10^7~\mathrm{Gpc^{-3}yr^{-1}}\,,
\end{equation}
comparable to the observed rate of \((7^{+9}_{-6})~\times~10^{7}\mathrm{~Gpc^{-3}~yr^{-1}}\) \citep{Bochenek_2020}.

The most significant source of uncertainty in this estimate is the average number density of ISOs \(n_\mathrm{ISO}\) around neutron stars.
In addition to the order-of-magnitude uncertainty in the number density of ISOs in the Solar neighbourhood, ISOs are not directly observable outside the Solar system so the density around different neutron stars across the local Universe may vary significantly from this value.
We expect the density of ISOs to vary across the Milky Way \citep{Hopkins_2023} and between different galaxies \citep{Williams_2022} due to a dependence of planetesimal formation on stellar metallicity \citep{Andama_2024}.
Secondly, the natal kick that a NS receives \citep{Hobbs_2005, Chatterjee_2005} can place them on an orbit removed from their galaxy's stars and ISOs, or completely eject them.
\cite{Sweeney_2022} model the Milky Way, predicting that 40\% of NS are completely ejected and the remaining NS occupy a disk with a vertical scale height five times that of the stellar disk.
ISOs are expected to retain the same spatial distribution as their parent stars \citep{Hopkins_2024}, therefore NS natal kicks could reduce the expected ISO-NS collision rate. 
This however will not be a major effect in elliptical galaxies, which make up \(\sim40\%\) of galaxies in the local universe brighter than \(M_r<-19\) \citep[][$M_r$ is the $r$-band absolute magnitude]{Lintott_2008}.

Finally, the timescale for a typical collision between an ISO and a neutron star (Eq. \ref{eq:gamma}) is far longer than that of repeating FRBs which occur on timescale of weeks.
Thus, we find that ISO-NS collisions cannot explain repeating FRBs and this mechanism cannot be the only source of FRBs\footnote{The contribution of multiple different mechanisms to the FRB population has been suggested in e.g. \cite{Gordon_2023}. Furthermore, \cite{Pleunis_2021} find differences in the emissions from repeaters and non-repeaters, implying repeaters and non-repeaters have different progenitors.}.
However, we can make testable predictions from the potential subset of FRBs caused by ISO-NS collisions.

\section{Emission Properties}\label{sec:properties}

We now explore the expected properties assuming the emission mechanism predicted in \cite{Geng_2015}, \cite{Dai_2016} and \cite{Dai_2020}.
When the tidal forces on the planetesimal exceed its tensile strength \(s\), it will be disrupted.
This has been observed in the Solar System, for example in the tidal disruption of comet Shoemaker-Levy 9 (SL9) by Jupiter \citep{Boss_1994}. 
Small bodies in the Solar System -- like asteroids and comets -- are rubble piles \citep[e.g.][]{Asphaug_1994, Walsh_2018} with a low tensile strength.
\cite{Colgate_1981} use a tensile strength \(s\sim10^9\mathrm{~Pa}\) and this value is continued in later works \citep[such as][]{Geng_2015, Dai_2016, Siraj_2019}.
However, asteroids and comets have significantly lower values, $s\sim 500\mathrm{~Pa}$ \citep[as constrained by spacecraft measurement and population analysis e.g.][]{Greenberg_1995, Attree_2018, Scheeres_2018}.
Likewise, the compressive strength, $P_0 \sim 100\mathrm{~MPa}$ \citep[e.g.][]{Jenniskens_2012, Flynn_2018, Pohl_2020}, is three orders of magnitude lower than in \cite{Colgate_1981}.
We use these current estimates of tensile and compressive strengths in the context of the \cite{Dai_2016} emission mechanism.

An ISO-NS collision produces homogenous tidal disruption, in contrast to a SL9-like disruption.
On a highly-elliptical orbit around Jupiter, SL9 passed close enough to the planet at perijove to partially disrupt into several fragments in 1992 \citep{Nakano_1993}.
However it then continued on its orbit, far enough from Jupiter that no more disruption events occurred, for an additional two years before colliding with the planet in 1994 \citep{Levy_1998}, allowing the fragments to drift a significant distance apart.
In an ISO-NS collision as modelled in \cite{Colgate_1981}, \cite{Geng_2015} and \cite{Dai_2016}, the planetesimal falls radially onto the NS, impacting the surface very shortly after fragmentation begins.
As it approaches the NS and the tidal forces on it continually increase the planetesimal will go through a rapid sequence of disruptions, with each generation of fragments undergoing their own fragmentation each time the remaining distance to the NS decreases by a factor of \(2^{-2/9}\simeq0.85\) \citep{Colgate_1981}.
Thus, the planetesimal will completely disrupt into a mostly-homogeneous stream of material, and we expect ISO-NS collisions to produce largely unstructured FRBs, without subbursts.\footnote{
This is the case even for contact binaries such as (486958) Arrokoth or 67P/Churyumov–Gerasimenko, made up of two distinct lobes held together by mutual gravity \citep{Scheeres_2007}, as the tidal force required to separate two lobes is the same order of magnitude as that required to disrupt each lobe individually.
}

\begin{figure}
    \includegraphics[width=\columnwidth]{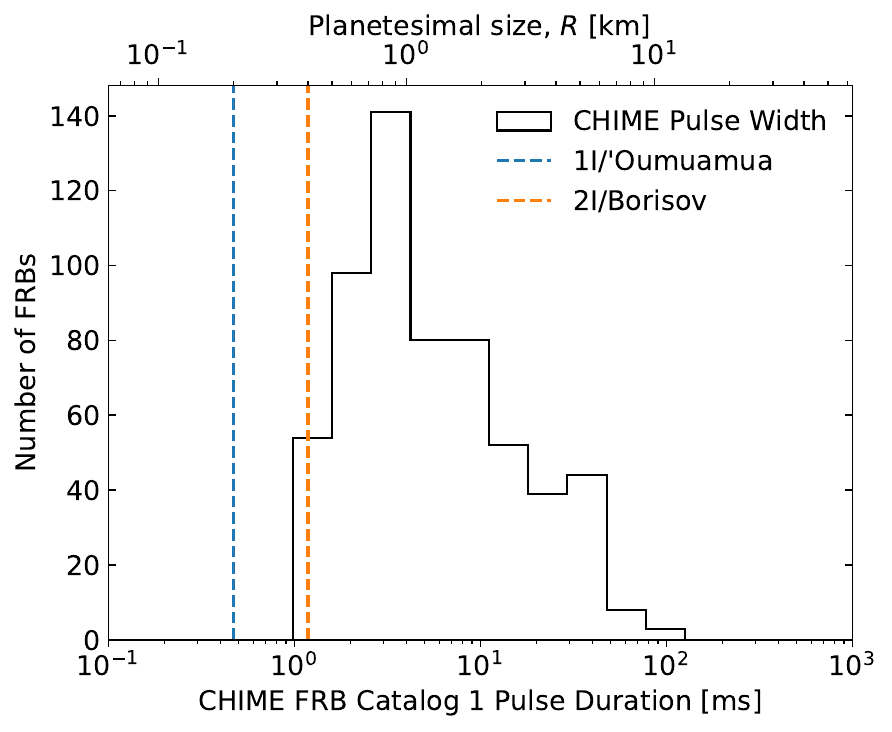}
    \caption{The CHIME Catalog 1 distribution of FRB boxcar pulse width.
    On the top axis, we show the equivalent planetesimal size, $R$, assuming tensile strength $s = 500 \mathrm{~Pa}$ and density $\rho_0 = 3\mathrm{~g~cm^{-3}}$.
    Dashed vertical lines show the estimated sizes of two observed ISOs, 1I/\okina Oumuamua and 2I/Borisov. The apparent sudden cutoff in FRB pulse duration at \(10^0\)~ms is due to CHIME's time resolution \citep[0.983 ms, ][]{CHIME/FRBCollaboration_2021}.
    \label{fig:frb_timescales}}
\end{figure}

After disruption, the time difference between the arrival of the leading and trailing edge of the disrupted planetesimal fragments is given by \cite{Colgate_1981}, with $\rho$ and $R$ as the density and radius of the planetesimal:
\begin{align}\label{eqn:Delta_t_R}
    \Delta t \simeq 4 \mathrm{~ms}
    \left(\frac{M_\mathrm{NS}}{1.4\mathrm{~M_\odot}}\right)^{-1/3} \left(\frac{\rho}{3\mathrm{~g~cm^{-3}}}\right)^{1/6} \nonumber  \\
    \cdot \left(\frac{s}{500\mathrm{~Pa}}\right)^{-1/6} \left(\frac{R}{1 \mathrm{~km}}\right)^{4/3}.
\end{align}
This relates planetesimal radius to the resulting FRB duration.
In Fig. \ref{fig:frb_timescales}, we show the histogram of FRBs' duration from CHIME Catalog 1 \citep{CHIME/FRBCollaboration_2021} with equivalent planetesimal radii, $R$.
Note that there is an apparent cutoff in FRB pulse duration at $\sim 1$ ms due to CHIME's time resolution \citep{CHIME/FRBCollaboration_2021}.
The distribution plotted here is subject to detection bias: as discussed later, the longer FRB pulses produced by larger ISOs are also brighter, and therefore more detectable.
Though a full model of the detection bias present in CHIME is beyond the scope of this work, we can qualitatively say that the distribution plotted in Fig~\ref{fig:frb_timescales} will be weighted towards longer pulse lengths and larger ISO radii than the underlying distribution.
However, the observed range of FRB durations corresponds to planetesimal radii \(400~\mathrm{m} \lesssim R \lesssim 10~\mathrm{km}\), still broadly consistent with the sizes of the two observed ISOs \citep{Jewitt_2017,Hui_2020}.

Using the updated value for tensile strength we can reinterpret the potential asteroidal cause of FRB 200428, first postulated in \cite{Dai_2020}.
FRB 200428 consists of two distinct subbursts of 0.6ms and 0.3ms separated by a large gap of 29ms \citep[See Fig.~1 of][]{CHIME/FRBCollaboration_2020}.
Assuming FRB 200428 was caused by an ISO-NS collision, \cite{Dai_2020} suggests that these subbursts were caused by two major fragments of a single body, split through tidal disruption during infall.
As discussed earlier in this section, we expect
instead that tidal disruption during infall onto a neutron star will not create two major fragments but instead a mostly-homogeneous stream of material.
Thus, we tentatively suggest instead that the large and complete separation of the sub-bursts is evidence that the colliding ISO was a binary: two separate bodies of radius 200m and 100m respectively, with a separation of several kilometres. 
Binaries are common within the Solar system \citep{Margot_2002, Walsh_2008, Fraser_2017}, but the potential existence of binary ISOs presents a fascinating opportunity for study.
If formed as a binary before ejection from their home planetary systems, the ejection mechanism must be gentle enough to allow the constituent bodies to remain gravitationally bound.
\cite{McDonald_2023} find that ejection by scattering off planets disrupts all binary asteroids; however, ejection mechanisms acting at larger stellocentric distances, such as stellar flybys \citep{Pfalzner_2021} or the effect of the Galactic tide on the outer edges of exo-Oort clouds \citep{Brasser_2010, Kaib_2011} may be sufficiently gentle.
Alternatively, binaries could perhaps be formed in the tidal disruption events that may be the source of some ISOs \citep{Walsh_2006, Cuk_2018, Zhang_2020}. 
An interesting prediction of connecting FRB subbursts to binary ISOs is that they should be largely limited to two subbursts, since triple (and higher) asteroids are much rarer. 
We find that of the non-repeating FRBs in the CHIME Catalog 1, there are 450 events consisting of one burst, 19 events with two subbursts, and only 5 events with more than two subbursts.
As discussed above ISO-NS collisions cannot be the source of all FRBs, but if they are the source of some, then FRBs may potentially be a probe for studying ISOs well beyond the limits of the Solar System.

Now we relate the luminosity distribution of FRBs to possible size distributions of ISOs.
Under the \cite{Dai_2016} mechanism most of the energy is released at frequency $\sim 1$ GHz, consistent with FRBs energy spectrum.
The luminosity is given by
\begin{align}
    L_\mathrm{tot} \sim 2.5 \times 10^{36} \mathrm{erg~s^{-1}} \left(\frac{P_0}{100\mathrm{~MPa}}\right)^{2/5}
    \left(\frac{s}{500\mathrm{~Pa}}\right)^{4/15} \nonumber\\ 
     \cdot  \left(\frac{R}{1\mathrm{~km}}\right)^{8/3}\left(\frac{\rho}{3\mathrm{~g~cm^{-3}}}\right)^{-2/3}
    \left(\frac{M_\mathrm{NS}}{1.4 \mathrm{M_\odot}}\right)^{19/12} \nonumber \\
     \cdot  \left(\frac{\mu_B}{10^{30}\mathrm{~G~cm^3}}\right)^{3/2}   \left(\frac{d}{10\mathrm{~km}}\right)^{-23/4} \left(\frac{\rho_c}{10^6\mathrm{~cm}}\right)^{-1}
\end{align}
where $\mu_B = B_\mathrm{surface} R_\mathrm{NS}^3$ is the magnetic dipole moment,
$B_\mathrm{surface}$ is the field strength at the NS surface,
$\rho_c$ is the curvature radius near the NS,
and $d$ is the distance from the ionized planetesimal fragments to the NS centre.
The strong dependence on $d$ implies that most of the energy is released near the NS surface at $d \simeq R_\mathrm{NS} \simeq 10\mathrm{~km}$.
This emission is strongly beamed by a factor \(f\sim1/100\), so the isotropic-equivalent energy emission is given by \(E_\mathrm{isotropic} \sim L_\mathrm{tot} \Delta t/f\). 
Observed FRBs have isotropic energy emission typically between $10^{35} - 10^{41}\mathrm{~erg}$ \citep{CHIME/FRBCollaboration_2020, Kirsten_2024}, equivalent to planetesimal sizes $0.5 - 10 \mathrm{~km}$ at typical pulsar magnetic field strengths, consistent with sizes inferred from FRB time duration. 

\begin{table}[]
    \centering
    \caption{Planetesimal size distributions power-law exponents \(q\) and the resultant predicted ISO-NS collision energy distribution exponent \(\gamma\), compared to the observed FRB energy distribution exponent.}
    \begin{tabular}{c | c | c}
        \toprule
        \(q\) & $\gamma$ & Description\\
        \hline \hline
        2.5 & 1.38 & Main asteroid belt \citep{Gladman_2009} \\
        2.8 & 1.45 & Streaming instability \citep{Simon_2016} \\
        3.5 & 1.63 & Faint TNOs \citep{Lawler_2018} \\
        \hline
         & $1.3^{+0.7}_{-0.4}$ & Observed FRBs \citep{Shin_2023} \\
        \hline
    \end{tabular}
    \label{tab:powerlaw}
\end{table}

The observed isotropic energy distribution of FRBs appears to follow a power law: 
\begin{equation}
    \dv{N}{E_\mathrm{isotropic} }
    \propto E_\mathrm{isotropic}^{-\gamma}
\end{equation}
with \(\gamma\) inferred to be $1.3^{+0.7}_{-0.4}$ by \cite{Shin_2023}.
Planetesimal size distributions are typically also treated as power laws or composite power laws: 
\begin{equation}
    \dv{N}{R} \propto R^{-q}
\end{equation}
with streaming instability modelling predicting $q \approx 2.8$ \citep{Simon_2016}, and observations of Solar system populations finding values of $q \approx 2.5$ for the main asteroid belt \citep{Gladman_2009} and $q \approx 3.5$ for faint trans-Neptunian objects \citep[e.g.][]{Lawler_2018}.
The ISO size distribution is completely unconstrained, but assuming ISOs have a similar power-law size distributions to Solar system populations we can relate the ISO power law exponent to the expected ISO-NS isotropic energy distribution exponent.
Since \(E_\mathrm{isotropic}\propto L_\mathrm{tot} \Delta t\propto R^4\), \(\gamma\) is related to \(q\) by \(\gamma=(3+q) / 4\).

Table \ref{tab:powerlaw} compares the expected values of $\gamma$ for ISO-NS collision isotropic energy distributions assuming different planetesimal size distributions to the observed FRB distribution.
Ideally we would compare our predictions for \(\gamma\) to the energy distribution of exclusively single pulses from non-repeating FRBs, as this is the type of FRB which our model predicts. 
However, the sample used by \cite{Shin_2023} includes repeating FRBs, albeit only the first detected burst from each, and use the combined energy released in the small number of FRBs with multiple pulses.
Since the majority of bursts are single pulses from non-repeaters we do not expect this to change the value of \(\gamma\) by a significant amount and find that the power-law scalings are within the uncertainty of observed FRB rates.

\section{Discussion}\label{sec:discussion}

Firstly, both NS and ISOs are long-lived and build up over time, meaning that the rate of ISO-NS collisions is decoupled from the current star formation rate.
This is in direct contrast to magnetars, a commonly-invoked FRB source \citep[e.g.][]{Metzger_2019, Wadiasingh_2019, Margalit_2019}, which have short lifetimes of 10~kyr \citep{Mondal_2021} meaning they are only present when and where star forming is actively occurring.
Several recent studies have found the redshift dependence of the FRB rate to be inconsistent with the evolution of the cosmic star formation rate \citep[e.g.][]{Hashimoto_2020a, Hashimoto_2022, Zhang_2022, Tang_2023, Lin_2024a, Lin_2024b,Zhang_2024,Chen_2024}, and FRBs cannot only originate from short-lived objects like magnetars. 
Instead of decreasing with cosmic time like the star formation rate \citep{Madau_2014} the FRB rate is increasing, as would be expected from long-lived progenitors.

Secondly, a small number of FRBs have been localised to their host galaxies, and though the majority of this sample are found in star-forming galaxies, those that originate in quiescent galaxies are unlikely to be from young progenitors \citep{Gordon_2023}.
If ISO-NS collisions are the source of a significant fraction of FRBs, we additionally would expect the reduced rate of ISO-NS collisions from disk galaxies as NS natal kicks cause a dependence of FRB rate on host galaxy morphology.

Therefore, we expect that insights into ISO-NS as a production mechanism will be constrained as more events are found and localised by SKA, CHIME, CHORD, and ASKAP \citep{Vanderlinde_2019, Hashimoto_2020b, Bhandari_2020, Michilli_2023}.

A final issue remains: the rate of observable ISO-NS collisions of \(\sim 10^7~\mathrm{Gpc^{-3}yr^{-1}}\) that we calculate in \S~\ref{sec:rate} is not directly comparable to the rate of FRBs of \((7^{+9}_{-6})~\times~10^{7}\mathrm{~Gpc^{-3}~yr^{-1}}\) inferred by \cite{Bochenek_2020}.
The rate we calculate corresponds to 1I-sized ISOs colliding with all NSs, whereas the observable FRB rate inferred by \cite{Bochenek_2020} corresponds to FRBs of at least the energy of FRB 200428, with an isotropic-equivalent energy release of \(2\times10^{35}\)~erg.
Ideally we would use a combination of the ISO size distribution and NS magnetic field distribution to calculate the rate distribution of ISO-NS collision energies, and compare the observable ISO-NS collision rate at the energy of FRB 200428.
However, given that the ISO size distribution is not well constrained, this is beyond the scope of this work.

\section{Conclusion}
The progenitors and emission mechanisms of FRBs are debated.
We revisit the emission mechanism of \cite{Geng_2015} and \cite{Dai_2016} which postulates FRBs are caused by collisions between neutron stars (NS) and planetesimals, and investigate the implications of these planetesimals being interstellar objects (ISOs).
As products of planetary formation, ISOs are common in the Milky Way and expected to be abundant across the Universe.
We have demonstrated that the expected ISO-NS collision rate is consistent with the cosmic rate of FRBs within the order-of-magnitude uncertainties.
Using updated planetesimal properties such as tensile strength, we have shown that observed FRB durations and energetics correspond to collisions with planetesimals of sizes in the range $0.4 - 10$ km, consistent with the two observed ISOs.
We have tentatively suggested FRBs with sub-bursts such as FRB 200428 could be caused by collisions with binary ISOs.
Finally, we have discussed the testable implications of ISO-NS collisions producing FRB-like signals on the redshift dependence of the FRB rate, and the morphology of galaxies which host FRBs.
Thus collisions between ISOs and NSs are feasible progenitors of one-off FRBs.

\begin{acknowledgements}
We thank the anonymous reviewer for their insightful comments.
We would like to thank Jocelyn Bell Burnell, Ayush Pandhi, and Alexander Andersson for insightful and helpful discussions.
MJH acknowledges support from the Science and Technology Facilities Council through grant ST/W507726/1.
\end{acknowledgements}

\bibliography{main}
\bibliographystyle{aasjournal}

\end{document}